\documentclass[prl,twocolumn,aps,showpacs,amsmath,amssymb]{revtex4}

\usepackage{graphicx}
\usepackage{dcolumn}
\usepackage{bm}

\begin{document}

\title{Ultrafast Quenching of Ferromagnetism in InMnAs \\ Induced
by Intense Laser Irradiation}

\author{J. Wang,$^{1,\dagger}$ C. Sun,$^1$ J.
Kono,$^{1,\ddagger}$ A. Oiwa,$^{2,\sharp}$ H. Munekata,$^3$
\L. Cywi{\'n}ski,$^4$}\author{L. J. Sham$^4$}
\affiliation{$^1$Department of Electrical and Computer
Engineering, Rice Quantum Institute, and Center for Nanoscale
Science and Technology, Rice University, Houston, Texas 77005,
U.S.A.\\ $^2$PRESTO, Japan Science and Technology Agency, 4-1-8
Honcho, Kawaguchi 332-0012, Japan\\
 $^3$Imaging Science and
Engineering Laboratory, Tokyo Institute of Technology, Yokohama,
Kanagawa 226-8503, Japan\\ $^4$Department of Physics, University
of California, San Diego, La Jolla, California 92093, U.S.A.}

\date{\today}

\begin{abstract}
Time-resolved magneto-optical Kerr spectroscopy of ferromagnetic
InMnAs reveals two distinct {\em demagnetization} processes
--- fast ($<1$ ps) and slow ($\sim$100 ps).  Both components
diminish with increasing temperature and are absent above
the Curie temperature.  The fast component rapidly grows with
pump power and saturates at high fluences ($>10$
mJ/cm$^2$); the saturation value indicates a {\em complete
quenching} of ferromagnetism on a sub-picosecond time scale. We
attribute this fast dynamics to spin heating through $p$-$d$
exchange interaction between photo-carriers and Mn ions while
the $\sim$100 ps component is interpreted as spin-lattice
relaxation.

\end{abstract}
\maketitle

There is much current interest in dynamical processes in
{\em magnetically-ordered} systems, both from scientific
and technological viewpoints \cite{ZhangetAl02book}. Pumping
a magnetic system with ultrashort laser pulses can strongly
alter the equilibrium among the constituents (carriers, spins,
and the lattice), triggering a variety of dynamical
processes.  Studying these processes can provide estimates for
the time scales / strengths of the various interactions.
Both metallic and insulating magnets have been studied,
exhibiting an array of new phenomena.
In particular, the discovery of {\em ultrafast demagnetization}
\cite{BeaurepaireetAl96PRL} suggested an ultrafast scheme for
magneto-optical data writing. At the same time, exactly how a
laser pulse can effectively change the magnetic moment is a
matter of debate
\cite{KoopmansetAl00PRL,GuidonietAl02PRL,ZhangetAl00PRL,Beaurepaire1etAl98PRB}.

Despite the large number of studies performed to date, the
microscopic understanding of effective energy transfer channels
among the subsystems is still elusive.  This is partially due to
the fact that, in the case of metallic ferromagnets, the
distinction between the ``carriers'' and ``spins'' is subtle, as the itinerant electrons contribute both to transport and magnetism.
On the other hand, antiferromagnetic insulators have
shown much slower dynamics, typically on the order of hundreds
of ps \cite{KimeletAl02PRL}, although femtosecond
demagnetization has been predicted \cite{GomezAbaletAl04PRL},
and ultrafast control of magnetism using off-resonance excitation
has been shown experimentally \cite{KimeletAl05Nature}.
Carrier-mediated ferromagnetism in (III,Mn)V semiconductors
provides an interesting alternative for this type of study.
Unlike in ferromagnetic metals, there is a clear distinction
between mobile carriers (holes) and localized spins (Mn ions),
and ferromagnetic order is realized through their strong
coupling ($p$-$d$ exchange interaction). This coupling in turn
makes the magnetic order sensitive to carrier density changes
via external perturbations
\cite{Koshiharaetal97PRL,Oiwaetal01APL,Oiwaetal02PRL,ChibaetAl03Nature}.

In this Letter we describe time-dependent magneto-optical studies
of laser-induced demagnetization dynamics in ferromagnetic
semiconductor InMnAs.  We observed a rapid ($<1$ ps) drop of
magnetization along with a slower demagnetization process on the
order of $\sim$100 ps. Both demagnetization components show
strong temperature and pump power dependence.  At high pump
fluences and under external magnetic fields, a complete
quenching of the ferromagnetic phase occurs, which appears as
a saturation of the photoinduced Kerr rotation with pump
fluence.  We attribute the observed fast and slow processes to
the manifestations of hole-localized spin and spin-lattice
interactions, respectively.

The main sample studied was an In$_{0.87}$Mn$_{0.13}$As(25
nm)/GaSb(820 nm) heterostructure with Curie temperature $\sim$60
K, grown by low-temperature molecular beam epitaxy on a
semi-insulating GaAs (100) substrate \cite{SlupinskietAl02JCG}.
Taking into account the electrical  and magnetic influences of
interstitial Mn ions which are  supposed to be double donors and
couple antiferromagnetically  with substitutional Mn ions
\cite{BlinowskietAl03PRB,WangetAl04JAP,EdmondsetAl05PRB}, the
hole density $p$ was extracted analytically from
the saturation magnetization data of the ferromagnetic
component. In the present sample, $p \simeq$ $4 \times 10^{20}$
cm$^{-3}$,
and the density of Mn spins participating in
ferromagnetic order is $\simeq 10^{21}$ cm$^{-3}$.

The two-color time-resolved magneto-optical Kerr effect
(MOKE) spectroscopy setup used consisted of an
optical parametric amplifier (OPA) pumped by a Ti:Sapphire-based
regenerative amplifier (Model CPA-2010, Clark-MXR, Inc.).
The OPA pump beam, which was linearly polarized and tuned to 2
$\mu$m, excited transient carriers near the band edge of InMnAs
and a very small fraction ($\sim$10$^{-5}$) of the regenerative
amplifier beam (775 nm) was used as a probe; the high photon
energy of the probe ensured to diminish the ``dichroic
bleaching'' effects \cite{KoopmansetAl00PRL} due to the pump
excited carriers.

\begin{figure}[floatfix]
\includegraphics [scale=0.6] {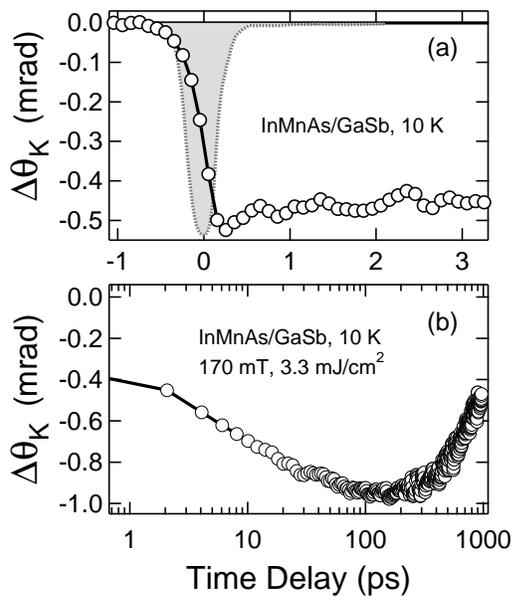}  
\caption{(a) The first 3 picoseconds of demagnetization dynamics
in InMnAs/GaSb.  Also shown is the cross-correlation between the
pump and probe pulses.  (b) Demagnetization dynamics convering
the entire time range of the experiment (up to $\sim$1 ns).
There is a slow demagnetization process, which follows the fast
component shown in (a) and completes only after $\sim$100 ps.}
\label{typical}\end{figure}

Typical data showing the general temporal profile of
the photoinduced Kerr angle change, $\Delta\theta_K$, is shown in
Fig.~\ref{typical}, for the first three ps [in (a)] and the
entire time range [in (b)].
The sign of $\Delta\theta_K$ is always negative, indicating
transient {\em demagnetization}. Distinct temporal
regimes can be identified: an initial
($<$ 1 ps) reduction in magnetization is followed by a slow
and gradual decrease (up to $\sim$100 ps), a
plateau region (up to $\sim$500 ps), and, finally, an
increase (i.e., recovery) toward the equilibrium value.  Also
shown in Fig.~\ref{typical} is the cross-correlation trace
between the pump and probe pulses, showing that the
initial ultrafast demagnetization is occuring even
faster than our time resolution ($\sim$220 fs).


\begin{figure}
\includegraphics [scale=0.6] {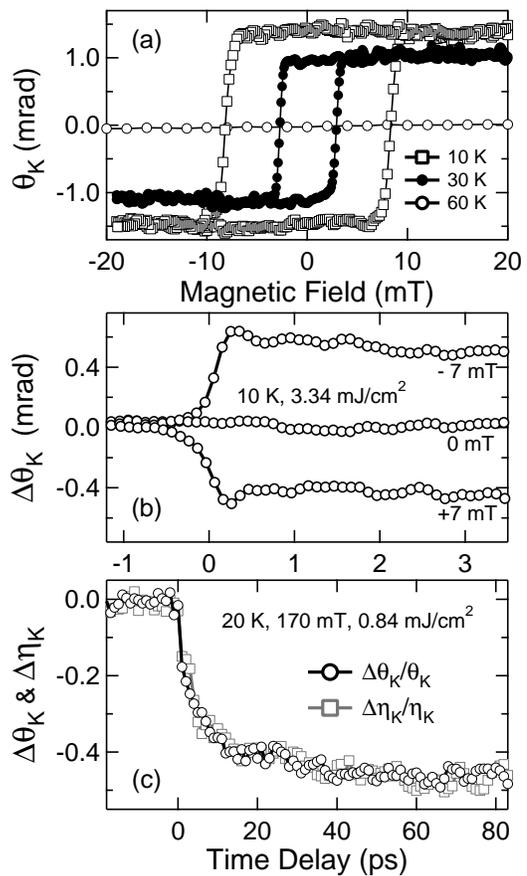}  
\caption{(a) Steady-state MOKE angle versus magnetic
field, taken with a 775-nm laser diode. (b)
Magnetic field dependence of ultrafast demagnetization dynamics.
 The sign of the MOKE angle change depends on the direction of
the applied magnetic field in a symmetric manner, as expected.
(c) Comparison between MOKE and MCD dynamics, showing no time
lag between them.} \label{mag-dep}\end{figure}

Figure \ref{mag-dep}(a) shows CW MOKE data taken with a 775 nm
laser diode.  The MOKE angle is $\sim$1 mrad and the coercivity
is $\sim$7.5 mT at 10 K.  MOKE dynamics in the first 3 ps is
shown in Fig.~\ref{mag-dep}(b) for different magnetic
fields: +7 mT, 0 mT, and $-$7 mT.  The sign of $\Delta\theta_K$ changes when the direction of the field is reversed, as expected for demagnetization.  In order to eliminate
the possibility that the induced MOKE changes are due to purely
optical effects, we also monitored both MOKE and reflection magnetic
circular dichroism (MCD) at the same time.  The exact
coincidence between MOKE and MCD, shown in
Fig.~\ref{mag-dep}(c), ensures that we are probing magnetic
properties (see, e.g., \cite{KoopmansetAl00PRL}).

\begin{figure}
\includegraphics [scale=0.5] {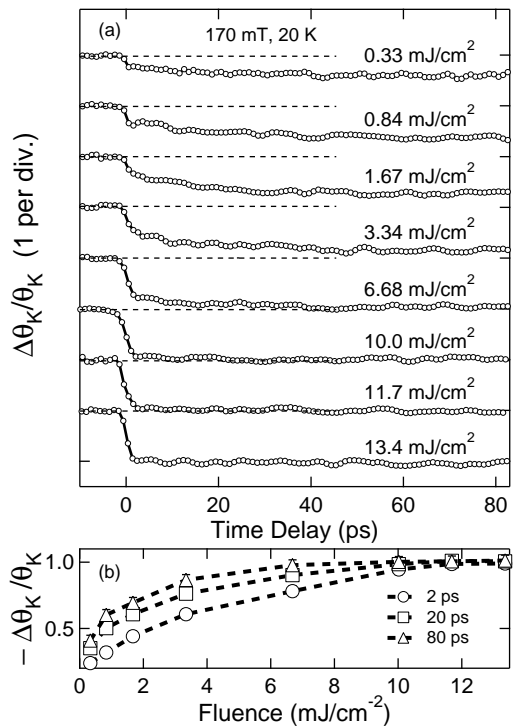}  
\caption{Normalized photo-induced MOKE angle change (a)
versus time for different pump fluences and (b) versus pump
fluence for different time delays.  At high fluences, the
signal saturates to $\sim$ 1, suggesting a complete quenching of
ferromagnetic order.} \label{power} \end{figure}

Photoinduced Kerr rotation dynamics, normalized by the MOKE angle
before the arrival of the pump, is plotted for different pump
fluences in Fig.~\ref{power}(a).  At low pump fluences, the fast
($<$ 1 ps) and slow ($\sim$100 ps) demagnetization components
co-exist.  However, as the fluence increases, the fast component
progressively becomes more dominant.  Around $\sim$10 mJ/cm$^2$,
there is no slow demagnetization process any more --- a sharp
initial drop is followed by a completely flat region.  This
``step-function'' like response remains stable both in shape and
magnitude against further increases in pump fluence to 13.4
mJ/cm$^2$, i.e., the demagnetization saturates.  The saturation
value is one, implying that the change in magnetization is
100\% [see also Fig.~\ref{power}(b)].
These results suggest that a {\em complete quenching} of the
ferromagnetic order is occurring on the order of several hundred
femtoseconds.


\begin{figure}
\includegraphics [scale=0.5] {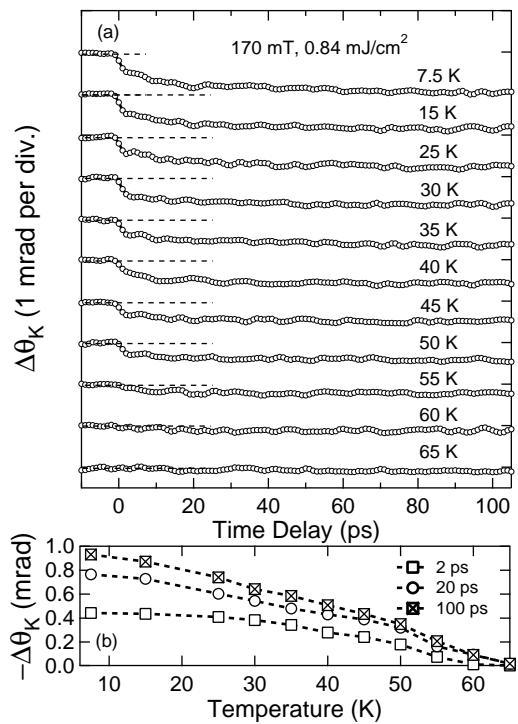}  
\caption{Photo-induced MOKE angle change (a)
versus time for different temperatures and (b)
versus temperature for different time delays.  The photo-induced
change is absent above the Curie temperature ($\sim$60 K).}
\label{temperature}\end{figure}

To substantiate the above claim, we performed temperature
dependent meaurements, and the resuts are shown in
Fig.~\ref{temperature}(a).  Here, a strong temperature dependence
of both the fast and slow components is seen. $\Delta\theta_K$
decreases drastically as the temperature approaches the critical
temperature $T_c$ ($\sim$ 60 K), and $\Delta\theta_K$ is {\em
absent above the $T_c$}. These facts further corroborate that the
transient Kerr rotation here measures the magnetization change of
the ferromagnetic state [Fig.~\ref{temperature}(b)].

In the standard picture of laser-induced demagnetization, the
heat deposited in the lattice is transferred into the spin
system through spin-lattice interaction (for microscopic model see e.g.  \cite{HubneretAl96PRB}).
The heating up of the spin system
then results in demagnetization, occurring at the time scale of
spin-lattice relaxation, $\tau_{sl}$, which is at least of the
order of 100 ps ($\approx 80$ ps in Gd
\cite{VaterlausetAl91PRL} and over ns in
paramagnetic CdMnTe \cite{StrutzetAl92PRL}). Much faster
demagnetization is expected when the process of direct
carrier-spin interaction is taken into account. Such an
interaction has been identified as the source of possible fast
demagnetization in metals \cite{BeaurepaireetAl96PRL}, and the
importance of the spin-orbit scattering of carriers has been
stressed \cite{Beaurepaire1etAl98PRB}.

Here, we can put these ideas on a stronger footing, exploiting
the simplicity of carrier-``spin'' coupling in diluted magnetic
semiconductors (DMS). The $p$-$d$ exchange interaction $H_{p-d}$
$\sim$$\beta\mathbf{S}\mathbf{s}$ (see, e.g.,
\cite{Furdyna88JAP}) couples the spins of delocalized holes with
localized Mn moments; the latter are the main source of the
macroscopic magnetization $\mathbf{M}$ (the carrier contribution
to $\mathbf{M}$ is very small). The mean-field part of this
interaction causes a spin-splitting of bands proportional to the
average value of Mn spin, which is exactly what is measured in
magneto-optical experiments. Photoexcitation creates a
nonequilibrium population of hot holes, with a blurred Fermi
surface. This strongly increases the number of states available
for spin-flip scattering, coming from the off-diagonal part of
$H_{p-d}$, allowing the flow of energy and angular momentum
between carriers and localized spins.  Demagnetization is caused
by this flow of polarization from Mn to holes, which is
sustained by efficient hole spin relaxation. The whole process
can be envisioned as the reverse of the Overhauser effect: the
excited carriers are becoming dynamically polarized at the
expense of the localized spins, and the dissipation of
magnetization occurs through spin relaxation in the carrier
system.

Spin-flip scattering has been used to describe the cw heating of
Mn by electrons in a paramagnetic DMS \cite{KonigetAl00PRB}, and the idea of
magnetization relaxation through spin-flips with carriers with
subsequent spin-relaxation by spin-orbit interaction can be
traced as far as the fifties \cite{Mitchell57PR} (recently
rejuvenated for DMS
\cite{TserkovnyaketAl04APL,SinovaetAl04PRB}).

At lowest pump fluences the demagnetization is $\sim20\%$
after 200 fs, and using the estimate of
$p \simeq 4 \cdot 10^{20}$ cm$^{-3}$ we see that every hole has to
participate in $\sim$1 spin-flip
($0.2\cdot{5 \over 2}\cdot10^{21}/(4\cdot10^{20}$))
to achieve such
an effect. This leads to an upper bound  for spin-relaxation time
$\tau_{s}\leq 200$ fs, which is reasonable: the momentum
scattering time in disordered III-V DMS is of the order of 10 fs
\cite{JungwirthetAl02APL}, and furthermore, in the excited
population other scattering mechanisms (carrier-carrier and
phonon scattering) play a significant role. Strong spin-orbit
interaction in the valence band leads to close correlation
between the momentum scattering and spin relaxation times.

In order to substantiate the above physical picture we used a single heavy-hole spin-split band to calculate the spin-flip scattering rates \cite{KonigetAl00PRB}: \begin{eqnarray*}
W_{m,m'} = \frac{2\pi}{\hbar}
\frac{\beta^{2}}{4}\sum_{\sigma}  \int
\frac{d^{3}k}{(2\pi)^{3}}  \int \frac{d^{3}q}{(2\pi)^{3}}
f_{\sigma}(\mathbf{k}) \Big( 1 - f_{-\sigma}(\mathbf{q}) \Big)
\\ \delta \Big( \epsilon_{-\sigma}(\mathbf{q}) -
\epsilon_{\sigma}(\mathbf{k}) + \delta \Big) | \langle m |
S_{\pm} | m' \rangle |^{2} \end{eqnarray*} where
$m=-\frac{5}{2},..., \frac{5}{2}$,
$\delta$ is the mean-field splitting of the Mn spin
levels, and $\sigma$ is the hole spin. The
occupation functions $f_{\sigma}(\mathbf{k})$ are thermal, with
a high effective temperature $T_{h}$ mimicking the highly
nonequilibrium state of holes, and with possible different Fermi
levels for two spin directions. The  dynamics of Mn spins is
described by a simple rate equation for diagonal elements of the
localized spin density matrix $ \dot{N}_{m} = \sum_{m'} (
W_{mm'}N_{m'} - W_{m'm}N_{m})$,
and the dynamics of the average hole spin $s$ is given by $ \dot{s} = -\gamma\dot{M} -
\frac{1}{\tau_{s}}(s-s_{eq}(M,T_{h}))$, where $M$ is the average Mn spin, $\gamma$ is the ratio of Mn to hole density, and $s_{eq}$ is the instantaneous equilibrium value of the average hole spin.

The results of calculations show a demagnetization of 10\%
within 200 fs for $T_{h}=1000$ K, $\tau_{s}=10$ fs,  and
$p=4\cdot 10^{20}$ cm$^{-3}$, which compares well with the
experiment. For much less favorable value of hole spin-relaxation time $\tau_{s}=100$ fs we get  $\Delta M/M \approx 5$ \%. More theoretical work on strong photoexcitation
of disordered III-V DMS and spin relaxation of hot holes is
needed to verify the parameters used above. However, it is clear
that even our simple model shows that ultrafast demagnetization
is possible in these materials.
The observed long-time dynamics is connected with the
``traditional'' pathway of heat transfer (with the intervention
of the lattice). The results are similar to those seen by
Kojima \cite{KojimaetAl03PRB} in GaMnAs, which were modeled by
the 3-temperature model
\cite{BeaurepaireetAl96PRL,HerouxetAl05unpub}. Our model
concentrates on the subpicosecond time-scale, where we
emphasize the role of polarization transfer between the systems
(with the possibility of ``bottleneck'' when $\tau_{s}$ is not
small enough), not only the heat exchange.

In conclusion, we have made the first observation of two distinct
demagnetization regimes in the dynamics of laser-excited InMnAs.
We interpret the novel demagnetization dynamics as a
result of ferromagnetic exchange couplings and spin-lattice
interactions, manifesting themselves as a fast ($<$1 ps)
and relatively slow ($\sim$100 ps) components, respectively.
The fast component completely quenches ferromagnetism at high
pump fluences. Systematic power and temperature dependence of
this ultrafast demagnetization provides convincing evidence
supporting the proposed physical picture.

\smallskip

This work was supported by DARPA (MDA972-00-1-0034), NSF
(DMR-0134058, DMR-0325474, DMR-0325599, INT-0221704), ONR
(N000140410657), and MEXT (No.14076210).

\medskip

\noindent$^{\dagger}$Present address: Lawrence Berkeley National
Laboratory, Berkeley, California.

\noindent$^{\ddagger}$To whom correspondence should be addressed.
Electronic address: kono@rice.edu.

\noindent$^{\sharp}$Present address: Department of Applied Physics,
University of Tokyo, 7-3-1 Hongo, Bunkyo-ku, Tokyo 113-8656,
Japan.



\end{document}